%% file: ms.tex
\documentclass[runningheads]{llncs}
\input{macros}

\begin{document}

\title{Compositional Reasoning for Side-effectful Iterators and Iterator Adapters}
% The long title just barely fits so no running title

\author{Aurel B\'il\'y\inst{1}\orcidID{0000-1111-2222-3333} \and
Jonas Hansen\inst{1}\orcidID{1111-2222-3333-4444} \and
Peter M\"uller\inst{1}\orcidID{2222--3333-4444-5555} \and
Alexander J. Summers\inst{2}\orcidID{2222--3333-4444-5555}}
\authorrunning{A. B\'il\'y et al.}
\institute{Department of Computer Science, ETH Zurich, Switzerland \and UBC, Canada}
\maketitle

\begin{abstract}
Iteration is a programming operation that traditionally refers to visiting the elements of a data structure in sequence. However, modern programming systems such as Rust, Java, and C\# generalise iteration far beyond the traditional use case. They allow iterators to be parameterised with (potentially side-effectful) closures and support the composition of iterators to form iterator \emph{chains}, where each iterator in the chain consumes values from its predecessor and produces values for its successor. Such generalisations pose three major challenges for modular specification and verification of iterators and the client code using them: (1)~How can parameterised iterators be specified modularly and their (accumulated) side effects reasoned about? (2)~How can the behaviour of an iterator chain be derived from the specifications of its component iterators? (3)~How can proofs about such iterators be automated?

We present the first methodology for the modular specification and verification of advanced iteration idioms with side-effectful computations. It addresses the three challenges above using a combination of inductive two-state invariants, higher-order closure contracts, and separation logic-like ownership. We implement and our methodology in a state-of-the-art SMT-based Rust verifier. Our evaluation shows that our methodology is sufficiently expressive to handle advanced and idiomatic iteration idioms and requires modest annotation overhead.

\keywords{Deductive verification \and Verified software \and Iterators \and Rust.}
\end{abstract}

% Main content begin

\section{Introduction}\label{sec:introduction}
  \input{sections/introduction.tex}

\section{Background}\label{sec:bg}
  \input{sections/bg.tex}

\section{Methodology}\label{sec:methodology}
  \input{sections/methodology.tex}

\section{Specifying individual iterators}\label{sec:iterators}
  \input{sections/iterators.tex}

\section{Specifying iterator chains}\label{sec:chains}
  \input{sections/chains.tex}

\section{Out-of-band contracts}\label{sec:impl}
  \input{sections/impl.tex}

\section{Soundness}\label{sec:soundness}
  \input{sections/soundness.tex}

\section{Evaluation}\label{sec:evaluation}
  \input{sections/evaluation.tex}

\section{Related work}\label{sec:rw}
  \input{sections/rw.tex}

\section{Conclusion}\label{sec:conclusion}
  \input{sections/conclusion.tex}

% Main content end

% \subsubsection{Acknowledgements} Please place your acknowledgments at the end of the paper, preceded by an unnumbered run-in heading (i.e. 3rd-level heading).

\clearpage

\bibliographystyle{splncs04}
\bibliography{paper}

%\appendix

%\section{Type-dependent contracts and coherence}\label{apx:typedef}
%  \input{sections/apx-typedep.tex}

\end{document}

%% file: macros.tex
\usepackage[T1]{fontenc}
\usepackage{graphicx}
\usepackage{amsmath}
\usepackage{amssymb}
\usepackage{booktabs}
\usepackage{tikz}
\usetikzlibrary{calc}
\usetikzlibrary{decorations.pathreplacing,calligraphy}
\usepackage{pgfplots}
\usepackage{siunitx}
\usepackage{listings, listings-rust}
\usepackage[normalem]{ulem}
\usepackage{adjustbox}
\usepackage{hyperref}
\usepackage{cleveref}
\usepackage{color}

\makeatletter
\newcommand*\codepoint[1]{\raisebox{1pt}[0.5em][0pt]{\tikz[baseline=(char.base)]{
  \node[
    fill=white,
    shape=circle,
    draw,
    inner sep=1pt,
    minimum width=10pt,
    scale=0.8,
  ] (realchar) {\textnormal{\textbf{\scriptsize\vphantom{WAH1g}#1}}};
  \node[
    inner sep=0pt,
    minimum width=0em,
  ] (char) {\phantom{\texttt{mmm}}};
}}}
\makeatother
\tikzset{
  state/.style={circle,fill=black,thick,inner sep=0pt,minimum size=0.15cm,align=center,font=\small},
  call/.style={rectangle,draw,thick,align=center,font=\small},
}

\lstnewenvironment{rust}{%
    \lstset{
      language=Rust,
      style=boxed,
      basicstyle=\fontsize{8}{10}\selectfont\ttfamily,
      mathescape=true,
      escapeinside={(*}{*)},
      tabsize=2,
    }
}{}
\newcommand{\code}[1]{\text{\lstinline[
  language=Rust,
  style=colouredRust,
  basicstyle=\small\ttfamily,
  mathescape=true,
  escapechar=~,
]`#1`}}

\newcommand*{\tableref}[1]{Table~\ref{table:#1}}
\newcommand*{\figref}[1]{Fig.~\ref{fig:#1}}
\newcommand*{\secref}[1]{Sec.~\ref{sec:#1}}

\newcommand{\eg}{{{e.g.\@}}}

\newcommand{\ie}{{{i.e.\@}}}

%% file: sections/introduction.tex
Iterators are a ubiquitous programming idiom traditionally used to abstractly and simply enumerate the elements of a collection. Verification of such simple use cases comes down to proving that an iterator yields all elements of a collection in a specified linear order and showing that the used iterator does not get invalidated \eg{} by concurrent modifications of the underlying collection~\cite{SAVCBSProceedings_2006}.

Modern programming languages such as C\#, Java, Python, and Rust support iteration patterns that go far beyond the traditional use case. For instance, iterators are used to perform computations over streams of values, such as computing a moving average; the computation itself is typically \emph{parameterisable} by custom code such as closures. In general-purpose languages, both the iterators themselves and these closures may perform side effects, \eg{} to modify the data structure in-place or to accumulate computations based on the values seen so far; code which uses such iterators depends on properties of this modified state.

Moreover, it is increasingly prevalent for languages to support the \emph{composition} of iterators, where an iterator processes values produced by another iterator (rather than obtained directly from a collection data structure). We follow the Rust terminology and call this kind of iterator an \emph{iterator adapter} (\emph{adapter} for short). Iterator adapters may be composed to form \emph{iterator chains}, where each iterator in the chain consumes the values from its predecessor in the chain and produces values for its successor; the overall chain acts as a composite iterator.

It is common for imperative languages to provide a variety of iterator adapters, \eg{} Java's Streams API, C\#'s LINQ, Python's iterables, and Rust's iterator adapters all support potentially-side-effectful variants of common functional programming operations such as \code{filter}, \code{fold}, and \code{map}. These languages also allow developers to implement their own custom iterators and iterator adapters.

The following \emph{Rust} example illustrates the use of iterators and adapters:

\begin{rust}
let mut sum = 0;
some_vector                     // start with a vector of integers
    .iter()                     // create an iterator over the vector
    .map(|x| { sum += x; sum }) // running totals of seen elements
    .filter(|x| x < 10)         // only keep totals smaller than 10
    .collect::<Vec<_>>();       // collect the results into a vector
\end{rust}

The function \code{iter} yields a traditional iterator for the underlying vector, which provides the input to the iterator chain. The \code{map} adapter is parameterised with a closure, which mutates the captured variable \code{sum}. The subsequent \code{filter} adapter, also parameterised with a closure for the filter criterion, removes elements, and finally \code{collect} stores the produced elements in another vector.

Such advanced iteration patterns lead to concise and readable code, but they also pose several challenges for modular specification and verification. Modularity is important to give correctness guarantees for libraries, to make verification scale, and to reduce the effort for re-verification when parts of a codebase change. The three main challenges are:

\begin{enumerate}
\item \emph{How to specify parameterised iterators modularly.}
The behaviour of iterators such as \code{map} depends on their argument closures. \emph{Modular} reasoning aims at verifying the iterator implementation against a specification that accounts for all possible closures that may be used by the client code. This requires a specification that is parametric in the argument closure, which is especially challenging for closures that have side effects such as the closure in line~4 of our example. The specification of the iterator must be strong enough (1)~to prove how the side effects affect the values produced by the iterator and (2)~to determine the \emph{accumulated} side effects of all calls to the iterator and its related closures by the time the iterator terminates. The latter is relevant because variables mutably captured by a closure, such as \code{sum} in the motivating example, become accessible to the client code once again when the iterator terminates (and, by extension, the closure expires).

In our example, assume that \code{some_vector} contains the values $6, 2, 9$. \code{map}'s specification must be strong enough to prove (1)~that the iterator produces the sums of the prefixes of the values it iterates over, that is, $6, 8, 17$, and (2)~that the final value of \code{sum} is $17$.

\item \emph{How to reason modularly about iterator chains.}
Modular verification requires that we can prove the behaviour of an iterator chain based on the specifications of its component iterators. In particular, verification of a chain must not require re-verification of its component iterators (whose code might not even be available); it should not be necessary to adapt the specification of an iterator for each and every instantiation and composition. Instead, the specification of an iterator must be agnostic to the source of values it consumes and to the downstream processing of the values it produces.

In our example, we want to prove that the final vector contains the values $6$ and $8$, without requiring the specifications of the intermediate adapters to either refer to the initial vector or depend on the subsequent adapters.

\item \emph{How to automate verification of modern iterator and adapter patterns.}
The behaviour of iterator chains depends on the behaviour of their component iterators, which may in turn depend on their argument closures. This hints at the need for higher-order logics (specifications which depend on specifications), which greatly complicate proof automation; we aim instead for proofs which can be automated with first-order SMT solvers.
\end{enumerate}

\subsubsection{State of the art.}

Prior work on iterator verification only partially addresses the challenges outlined above. Pereira~\cite{Pereira_Iterators_2018} defines a modular verification technique using two predicates per iterator to specify the sequence of produced values and a termination condition. This technique is automated using Why3~\cite{Filliatre_Why3_2013}. It partially addresses Challenge~1 by providing abstract, implementation-independent specifications for individual iterators. However, although higher-order iterators are discussed, this only refers to simple iterators parameterised by side-effect-free functions, as in a functional \code{fold} operation. Support for iterators with side effects (for instance, to prove the final value of \code{sum} in our example) is discussed as future work in the form of a tool which rewrites client code into equivalent loops, leaving the appropriate specification and verification to manual proof effort. Iterator composition is not addressed.

Verification of higher-order functions such as \code{fold} in Rust is partially addressed in existing work on \emph{closure} verification~\cite{Wolff_Closures_2021}, allowing one to reason modularly about side-effectful closures, such as the argument to \code{map} in our example. However, this technique does not provide specification or verification support for iterators and adapters that \emph{use} closures in well-known but functionally complex ways (\eg{} as an argument to a \code{fold}).

\subsubsection{This work.}

We present the first modular specification and verification technique for advanced iteration patterns that addresses the three challenges above. To address Challenge~1, we associate iterators with four key specifications: (1)~the sequence of elements, \code{produced}, returned by the iterator so far, (2)~a \code{completed} predicate that expresses when iteration terminates, (3)~a two-state postcondition \code{step} that relates two consecutive iterator states and (4)~a two-state predicate \code{leadsto} that inductively relates \emph{any} previous state of the iterator to \emph{any} more recent state. The first two points are inspired by Pereira~\cite{Pereira_Iterators_2018}, but importantly our technique can also handle \emph{side-effectful} iterators by characterising the intermediate \emph{states} occurring during an iteration, not only the returned elements. We employ so-called \emph{call descriptions}~\cite{Wolff_Closures_2021} to parameterise the \code{step} and \code{leadsto} predicates with the behaviour of closures (\eg{} for the \code{map} and \code{filter} iterators above). In order to reason about the \emph{accumulated} effects of an iterator, we prove that each call to \code{next} satisfies the transitive two-state invariant \code{leadsto}. Consequently, this invariant also holds between the initial and the final state of the entire iteration and, thus, is able to characterise its overall effects.

We show that our newly developed methodology generalises directly to iterator chains (Challenge~2): the \code{leadsto} predicate of an iterator adapter may refer to the \code{leadsto} predicate and the \code{produced} sequence of its predecessor in an iteration chain to specify the successor's behaviour in terms of the behaviour of the predecessor. In particular, these specification ingredients let the successor maintain ghost data structures that keep track of all states of the predecessor iterator and of any argument closures that occur during the iteration. We can then reason about side effects via invariants over these ghost structures.

Our methodology builds exclusively on specification constructs for which automated verification techniques exist (ghost fields, invariants, and closure call descriptions). It can, thus, be automated in SMT-based verifiers, as we demonstrate through an extension of the Rust verifier Prusti~\cite{Prusti_2019} (Challenge~3).

We present our technique in the context of Rust, whose ownership type system complements our methodology by preventing concurrent modifications of an iterated-over data structure, or undesirable interference between iterators. However, our technique would apply equally to other languages if augmented with an alternative ownership-like technique such as separation logic~\cite{Reynolds_SL_2002}.

\subsubsection{Contributions and outline.}

The main contributions of our paper are:

\begin{itemize}
  \item We present the first specification and verification methodology for general side-effectful iterators ($\rightarrow$ \secref{methodology}, \secref{iterators}).

  \item We demonstrate how to use this methodology to reason about the effects and resulting values of complex iterator chains ($\rightarrow$ \secref{chains}).

  \item We show how to express these specifications modularly for existing iterator hierarchies ($\rightarrow$ \secref{impl}).

  \item We implement our work in the Prusti verifier~\cite{Prusti_2019} and demonstrate its expressiveness on several challenging examples ($\rightarrow$ \secref{evaluation}).
\end{itemize}

\noindent
Additionally, we provide the necessary background on Rust ($\rightarrow$ \secref{bg}), explain why our approach is sound ($\rightarrow$ \secref{soundness}), and discuss related work ($\rightarrow$ \secref{rw}).

%% file: sections/bg.tex
In this section we briefly describe Rust, focusing on its ownership type system ($\rightarrow$ \secref{bg-rust}), then we discuss the relevance of this type system to verification in general and our methodology in particular ($\rightarrow$ \secref{bg-rust-ver}).

\subsection{Rust}\label{sec:bg-rust}

Rust~\cite{Rust_2014} is a modern systems programming language that is rapidly gaining popularity. Among mainstream languages, it is unique due to its \emph{ownership} type system, which eliminates certain categories of common software bugs, such as data races, use-after-free bugs, dangling pointers, and null pointer dereferences.

The Rust type system maintains a crucial invariant: any value is either mutable or shared, but never both.
To maintain this invariant, values are categorised as: \emph{owned}, \emph{mutably borrowed}, or \emph{immutably borrowed}. All values have a single \emph{owner}, which by default has exclusive (read/write) capabilities on the data. Since ownership is exclusive, no other thread or function can access the owned value without \emph{borrowing it} via a reference. A value can be \emph{mutably borrowed} to create exactly one mutable reference. While the mutable reference exists, it has the unique capability to read from and modify the borrowed value. Immutable, or \emph{shared} references, on the other hand, have a \emph{read-only} capability for the borrowed value. Unlike mutable references, a value can be immutably borrowed any number of times. When all borrows expire (go out of scope) the borrowed-from value becomes writable again. The following snippet demonstrates some of these concepts; for a full overview of Rust we refer the reader to the Rust Book~\cite{RustBook_2018}.

\begin{rust}
struct Point { (*\codepoint{A}*)
    x: i32, y: i32,
}
fn origin() -> Point { (*\codepoint{B}*)
    Point { x: 0, y: 0 }
}
fn get_point_x(pt: &Point) -> &i32 { (*\codepoint{C}*)
    &pt.x
}
fn set_point_x(pt: &mut Point, x: i32) { (*\codepoint{D}*)
    pt.x = x;
}
\end{rust}

The above code declares a 2D point structure \codepoint{A} consisting of two integer coordinates. The \code{origin} method \codepoint{B} returns an owned \code{Point} to the caller with both coordinates at zero (types not prefixed with an \code{&} are owned). \code{get_point_x} \codepoint{C} takes a shared reference to a point (indicated by \code{&Point}), and returns a shared reference to its \code{x} coordinate. Finally, \code{set_point_x} \codepoint{D} takes a mutable reference to a point (indicated by \code{&mut Point}), and sets its \code{x} coordinate to the given value.
The following code makes use of these methods:

\begin{rust}
let mut pt = origin();
set_point_x(&mut pt, 5);
let x = get_point_x(&pt);
// set_point_x(&mut pt, 6); // not allowed!
println!("x coordinate is: {}", *x);
set_point_x(&mut pt, 7);
\end{rust}

The variable \code{pt} is an owned \code{Point} instance. The first call to \code{set_point_x} is made with a mutable reference to \code{pt}. The \code{x} variable creates a shared borrow into \code{pt}. The mutable borrow in the call to \code{set_point_x} (on line 5) is rejected by the type system because it would create a mutable reference to a \code{Point} object while an immutable reference to its \code{x} field still exists.

The Rust type system supports concrete types (such as structs, enums, and unions) and \emph{traits}. Traits make it possible to define an interface shared across many types, optionally including default implementations for its methods. Traits are also used to declare the iterator interface in Rust, as we will see  in~\secref{iterators-specext}.

\subsection{Deductive verification in a Rust setting}\label{sec:bg-rust-ver}

The notion of ownership in Rust's type system is closely connected to ownership in separation logic (SL)~\cite{Reynolds_SL_2002}. Earlier work on the Prusti verifier~\cite{Prusti_2019} exploits this correspondence to automatically extract memory safety proofs in SL from Rust type and borrow information. This \emph{core proof} contains all information to enable \emph{framing}, that is, proving that a given heap property is not affected by a heap modification. As a result, user-specified functional annotations can be proved easily by conjoining them to the assertions of the core proof. Prusti verifies the combined program using the Viper verification framework~\cite{Viper_2016}.

Our verification methodology for iterators leverages Rust's ownership system in two ways. First, ownership defines the heap fragment that an iterator and its argument closures may access and modify, and prevents concurrent modifications. Second, we can focus on the essential functional properties of iterators and rely on Rust's type system for memory safety and framing, analogously to Prusti. Nevertheless, we are confident that our methodology can be combined with SL to handle languages without a built-in ownership system.

%% file: sections/methodology.tex
\newcommand{\sq}[1]{$#1_0, #1_1, \dots$}

In this section we introduce a general-purpose methodology for reasoning about iterators. We describe a model of iterators ($\rightarrow$~\secref{methodology-model}), introduce four predicates that allow one to specify the behaviour of iterators ($\rightarrow$~\secref{methodology-components}), and present the proof obligations needed to verify that an iterator implementation satisfies this specification ($\rightarrow$~\secref{methodology-next}). We illustrate our methodology on two examples ($\rightarrow$~\secref{methodology-examples}). Finally, we summarise how our methodology addresses the challenges outlined in the introduction ($\rightarrow$~\secref{methodology-challenges}).

\subsection{Iteration model}\label{sec:methodology-model}

An \emph{iterator} can be queried repeatedly for values using a \code{next} method, producing a sequence of values (\sq{v} in the diagrams below). Calls to the \code{next} method change the internal state of the iterator (\sq{I} below).

\input{figures/dia1.tex}

The sequence of produced values can be finite or infinite. In the former case, \code{next} yields a designated value $\bot$ to signal that the iteration has completed\footnote{In Rust, this corresponds to an \code{Option::None}.}. We assume that once an iterator has completed, it will remain completed, which is the case in many programming languages. In Rust, this assumption corresponds to a so-called \emph{fused} iterator (\code{FusedIterator} trait), and it is satisfied by the vast majority of Rust's standard library iterators.

\emph{Iterator adapters} are iterators which use the results of a previous iterator to produce their values. For instance, the \code{Map} adapter applies a function \code{f} to the outputs \sq{v} of the previous iterator to produce its result values \sq{m}. The function \code{f} may itself be a closure with potentially mutable captured variables, which means it has its own state (\sq{F}):

\input{figures/dia3.tex}

\input{figures/dia45.tex}

The overall state of the \code{Map} adapter $M_i$ comprises (and encapsulates) the state of the previous iterator $I_i$ and that of its argument closure $F_i$. \code{Map}'s \code{next} method internally manipulates both state components as shown in \figref{map-filter} (left).

Each call to \code{Map}'s \code{next} method triggers exactly one call to the predecessor's \code{next} method. Other adapters have more complex behaviours. For instance, the \code{Filter} (\figref{map-filter} (right)) adapter calls its predecessor's \code{next} method until the provided value satisfies the filter criterion determined by an argument closure (or the predecessor has completed) A specification and verification methodology for advanced iterators must be able to capture the evolution of such composite iterator states and the interactions between different iterators.

\subsection{Specification components}\label{sec:methodology-components}

Our methodology uses four specification components to specify the states of and values returned by iterators. We introduce them here using a mathematical notation and show a concrete syntax in Rust later, in~\secref{iterators-specext}.

As in prior work~\cite{Pereira_Iterators_2018}, we associate each iterator with two ghost (specification-only) functions to specify the \emph{values} returned by the iterator. Function \code{produced} yields the sequence of elements returned \emph{so far}:  $\code{produced($I_k$)} = \left[ v_0, v_1, \dots, v_{k - 1} \right]$. Function \code{completed} yields true iff the iterator has completed. In other words, a call to \code{next} with the initial state $I_k$ returns $\bot$ iff \code{completed($I_k$)} holds.

The evolution of an iterator's \emph{state} across a single call to \code{next} is characterised using the third component of our methodology: a two-state predicate \code{step}. For a call to \code{next} with the initial iterator state $I_k$, the updated iterator state $I_{k + 1}$ and the returned value $v_k$, \code{step($I_k$, $I_{k + 1}$, $v_k$)} must hold.

To reason about the \emph{accumulated} effects of an iterator, in general concerns an \emph{unbounded} number of \code{next} calls. This motivates our fourth key component: we associate each iterator with a predicate \code{leadsto} that represents an \emph{inductive}, \emph{two-state} invariant. This invariant relates the current iterator state to \emph{any previous} iterator state. It represents the reflexive, transitive closure of the two-state postcondition \code{step} (which relates \emph{consecutive} iterator states): $\code{leadsto($I_k$, $I_l$)} \Leftrightarrow (\forall i \cdot k \leq i < l \Rightarrow \code{step($I_i$, $I_{i + 1}$, $\_$)})$ for any $0 \leq k \leq l$.

\input{figures/dia78.tex}

When an iterator is used in a loop, we can naturally make use of \code{leadsto} in the loop invariant: \code{leadsto($I_0$, $I_k$)} must hold at every iteration, where $I_0$ is the state of the iterator before the loop, and $I_k$ is the current state.

\subsection{Proof obligations}\label{sec:methodology-next}

To use our methodology, programmers need to define the four specification components described in the previous subsection for each concrete iterator implementation. Our methodology then imposes the following proof obligations. First, we check that \code{leadsto} includes the reflexive, transitive closure of \code{step} (see previous subsection). If this well-formedness check fails, the program is rejected. Second, we check whether the definitions of the four components correctly reflect the behaviour of the iterator implementation. This can be done by verifying that the implementation of the \code{next} method satisfies the following five postconditions. In these conditions, $I_k$, $I_{k + 1}$, and $v_k$ denote the prestate, poststate, and result value of \code{next}, respectively. For simplicity we assume that the iterator has no methods that modify its state, other than \code{next}.

\begin{itemize}
  \item $Q_1 \equiv \code{step($I_k$, $I_{k + 1}$, $v_k$)}$:
\code{step} reflects the behaviour of a single call to \code{next}.

  \item $Q_2 \equiv \code{completed($I_k$)} \Leftrightarrow v_k = \bot$: \code{completed} reflects correctly whether the iterator returns a value.

  \item $Q_3 \equiv \code{completed($I_k$)} \Rightarrow \code{produced($I_{k + 1}$)} = \code{produced($I_k$)}$: if the iterator has completed, the \code{produced} sequence is left unchanged.

  \item $Q_4 \equiv \neg \code{completed($I_k$)} \Rightarrow  \code{produced($I_{k + 1}$)} = \code{produced($I_k$)} \code{++} [ v_k ]$: if the iterator has \emph{not} completed, the \code{produced} sequence is extended by the returned value.

  \item $Q_5 \equiv \code{completed($I_k$)} \Rightarrow \code{completed($I_{k + 1}$)}$: iterator completion is monotonic.
\end{itemize}

As usual, these postconditions may be assumed after each call to \code{next}, which allows client code to reason about the results and effects of an iterator. We show examples in the next subsection.

\subsection{Examples}\label{sec:methodology-examples}

In this subsection, we show how to instantiate our specification components for an iterator and an iterator adapter. In both examples, we assume that the concrete iterator implementation maintains a ghost field \code{p} that contains the elements produced so far; \code{produced} is then defined to simply return \code{p}. Proof obligations $Q_3$ and $Q_4$ above ensure that the \code{next} method updates \code{p} correctly. In the definitions below, we still use mathematical notation over named states (as in our diagrams). We will define a suitable concrete specification syntax in~\secref{iterators}.

\subsubsection{Counter iterator.}

A counter iterator that yields the numbers between 0 and \code{bound} can be (partially) specified as follows. Here, \code{ctr} is the field of the iterator that stores the next number to return.

\begin{align*}
  \forall 0 \leq k \leq l \; \cdot && \code{completed($C_k$)} \triangleq\; &\code{$C_k$.ctr} = \code{$C_k$.bound} \\
  && \code{step($C_k$, $C_{k + 1}$, $v_k$)} \triangleq\;
    &v_k = \code{$C_k$.ctr} \land \code{$C_k$.ctr} + 1 = \code{$C_{k + 1}$.ctr} \\
  && \code{leadsto($C_k$, $C_l$)} \triangleq\;
    &\code{$C_k$.ctr} \leq \code{$C_l$.ctr}
\end{align*}

The \code{leadsto} definition declares that the counter is monotonically increasing. Thus, even across a statically unknown number of calls (\eg{} across a loop), we always know that the counter will not emit values lower than ones we have previously observed.

This specification omits various details about the returned values. To obtain a more comprehensive specification, \code{step} and \code{leadsto} need to constrain \code{produced}, for instance to express that \code{produced[$i$]} contains the value $i$. We omit the details for simplicity, but our methodology can easily express them when desired.

\subsubsection{Map iterator adapter.}

As explained, a \code{Map} iterator adapter produces a value by calling a closure on the next value from the iterator it adapts. We specify this as follows: (1)~The \code{Map} iterator has completed iff the input iterator has completed. (2)~Each execution of \code{next} performs a step on the input iterator and calls the map's closure on the returned value. (3)~The state of the \code{Map} iterator evolves according to the specifications of the input iterator and the closure.

The following definitions reflect this intuition. As before, $M_i$ is the state of the entire \code{Map} iterator, consisting of the state of the input iterator $I_i$ and of the closure $F_i$. Note that these states include the identities of those sub-objects; that is, $F_i$ includes the information of which closure to call, along with the values of any captured variables. In the definition of \code{step}, we use a \emph{call description}, which has been proposed by Wolff et al.~\cite{Wolff_Closures_2021} to specify the invocation of closures. $(F_k, v_k) \rightsquigarrow (F_{k+1}, m_k)$ expresses that one call to the closure with state $F_k$ is made with argument $v_k$, resulting in an updated closure state $F_{k+1}$ and return value $m_k$. The changes to the closure state are constrained by the closure's \emph{history invariant} \code{hist_inv}, a two-state invariant relating any previous closure state to any new closure state. We use this history invariant in the definition of \code{leadsto}. We will describe call descriptions in more detail in~\secref{iterators-syntax}.

\begin{align*}
  \forall 0 \leq k \leq l \; \cdot && \code{completed($M_k$)} \triangleq\; &\code{completed($I_k$)} \\
  && \code{step($M_k$, $M_{k + 1}$, $m_k$)} \triangleq\;
    &\exists v_k \cdot \code{step($I_k$, $I_{k + 1}$, $v_k$)} \land \\
  && &\land (F_k, v_k) \rightsquigarrow (F_{k+1}, m_k) \\
  && \code{leadsto($M_k$, $M_l$)} \triangleq\; &\code{leadsto($I_k$, $I_l$)} \land \code{hist_inv($F_k$, $F_l$)}
\end{align*}

A client that knows the concrete input iterator and closure can combine this knowledge with this specification together with the postcondition of \code{next} to determine the values returned by the \code{Map} iterator.

\subsection{Challenges revisited}\label{sec:methodology-challenges}

The introduction presented three main challenges for the specification and verification of advanced iterators. In this subsection we summarise how our novel methodology addresses each of them.

\emph{How to specify parameterised iterators modularly.} We capture the result and effects of a single call to \code{next} using the two-state \code{step} predicate. When the iterator is parameterised with a closure (as in the \code{Map} example), we use the call descriptions from earlier work~\cite{Wolff_Closures_2021} as a means of abstractly and generically describing the results and effects of calls to the closure.

We capture the \emph{accumulated} results and effects of an iteration using the \code{produced} sequence and the reflexive, transitive \code{leadsto} predicate. When the iterator state includes closures, we use two-state invariants on the \emph{closure} state (again following Wolff et~al.) to express how they evolve during an iteration.

Postconditions for the \code{next} method tie together these predicates with the concrete implementation of the iterator. They allow clients to determine the behaviour of a single call to \code{next} (using \code{step}) as well as of a full iteration (using \code{leadsto}). In particular, clients, which are aware of the concrete argument closure passed to the iterator, can use these postconditions to determine both the sequence of returned values and all relevant accumulated side-effects (\eg{} on closure-captured state).

\emph{How to reason modularly about iterator chains.} To handle iterator chains, the specification of iterator adapters may refer to the \code{step} and \code{leadsto} predicates of the previous iterator in the chain. This makes the specification of the adapter \emph{parametric} in the behaviour of the input iterator. Clients, which are aware of the concrete iterator used as input to the adapter, know how to resolve this parameterisation to obtain the concrete behaviour of the adapter. In particular, neither the input iterator nor the adapter need to be re-verified when forming an iterator chain.

\emph{How to automate verification of modern iterator and adapter patterns?} As noted in~\secref{introduction}, this challenge is addressed by encoding the methodology into first-order logic components suitable for verification with SMT-based verifiers. The four predicates defined above are encoded as uninterpreted functions, their argument states are encoded using snapshots (explained in~\secref{iterators-snapshots}). A suitable first-order encoding of call descriptions is provided in prior work~\cite{Wolff_Closures_2021}.

%% file: figures/dia1.tex
\begin{center}\begin{tikzpicture}
  \node[state,label={$I_0$}] (it0) at (0, 0) {};
  \node[state,label={$I_1$}] (it1) at (2, 0) {};
  \node[state,label={$I_2$}] (it2) at (4, 0) {};
  \node                      (itt) at (6, 0) {\dots};

  \node[state,label={$v_0$}] (vl0) at (2.5, -.8) {};
  \node[state,label={$v_1$}] (vl1) at (4.5, -.8) {};
  \node                      (vlt) at (6.5, -.8) {\dots};

  \node[call] (next0) at (1, 0) {\code{next}};
  \node[call] (next1) at (3, 0) {\code{next}};
  \node[call] (next2) at (5, 0) {\code{next}};

  \draw[->] (it0)   -- (next0);
  \draw[->] (next0) -- (it1); \draw[->] (next0.345) -- (vl0);
  \draw[->] (it1)   -- (next1);
  \draw[->] (next1) -- (it2); \draw[->] (next1.345) -- (vl1);
  \draw[->] (it2)   -- (next2);
  \draw[->] (next2) -- (itt); \draw[->] (next2.345) -- (vlt);
\end{tikzpicture}\end{center}

%% file: figures/dia3.tex
\begin{center}\begin{tikzpicture}
  \node[state,label={$I_0$}] (it0) at (0,  0) {};
  \node[state,label={$I_1$}] (it1) at (4,  0) {};
  \node[state,label={$I_2$}] (it2) at (8,  0) {};
  \node                      (itt) at (10, 0) {\dots};

  \node[state,label={$F_0$}] (fn0) at (0,  -1) {};
  \node[state,label={$F_1$}] (fn1) at (4,  -1) {};
  \node[state,label={$F_2$}] (fn2) at (8,  -1) {};
  \node                      (fnt) at (10, -1) {\dots};

  \node[state,label={$v_0$}] (vl0) at (2,    -.5) {};
  \node[state,label={$v_1$}] (vl1) at (6,    -.5) {};
  \node                      (vlt) at (10.5, -.5) {\dots};

  \node[state,label={$m_0$}] (ml0) at (4.5, -1.5) {};
  \node[state,label={$m_1$}] (ml1) at (8.5, -1.5) {};

  \node[call] (next0) at (1, 0) {\code{next}};
  \node[call] (next1) at (5, 0) {\code{next}};
  \node[call] (next2) at (9, 0) {\code{next}};

  \node[call] (call0) at (3, -1) {\code{call}};
  \node[call] (call1) at (7, -1) {\code{call}};

  \draw[->] (it0)   -- (next0);
  \draw[->] (next0) -- (it1); \draw[->] (next0.345) -- (vl0);
  \draw[->] (it1)   -- (next1);
  \draw[->] (next1) -- (it2); \draw[->] (next1.345) -- (vl1);
  \draw[->] (it2)   -- (next2);
  \draw[->] (next2) -- (itt); \draw[->] (next2.345) -- (vlt);

  \draw[->] (fn0)   -- (call0); \draw[->] (vl0) -- (call0.165);
  \draw[->] (call0) -- (fn1); \draw[->] (call0.345) -- (ml0);
  \draw[->] (fn1)   -- (call1); \draw[->] (vl1) -- (call1.165);
  \draw[->] (call1) -- (fn2); \draw[->] (call1.345) -- (ml1);
  \draw[->] (fn2)   -- (fnt);
\end{tikzpicture}\end{center}

%% file: figures/dia45.tex
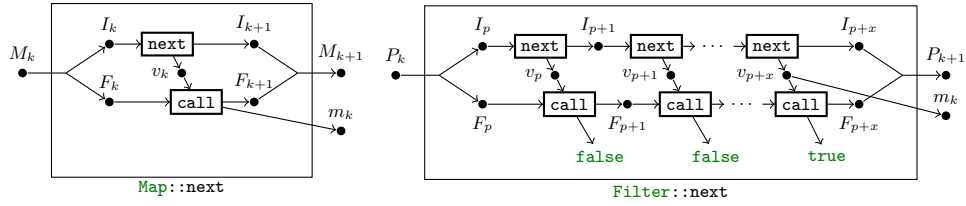
\begin{figure}[t]
\begin{adjustbox}{center}
\begin{tikzpicture}[scale=0.77, every node/.style={transform shape}]
  \node[state,label={$M_k$}]     (mak)  at (0.5, -.5) {};
  \node[state,label={$M_{k+1}$}] (mak1) at (6, -.5) {};

  \coordinate (join1) at (1.25, -.5) {};
  \coordinate (join2) at (5.25, -.5) {};

  \node[state,label={$I_k$}]     (itk)  at (2, 0) {};
  \node[state,label={$I_{k+1}$}] (itk1) at (4.5, 0) {};

  \node[state,label={$F_k$}]     (fnk)  at (2, -1) {};
  \node[state,label={$F_{k+1}$}] (fnk1) at (4.5, -1) {};

  \node[state,label={180:$v_k$}] (vlk) at (3.25, -.5) {};

  \node[state,label={$m_k$}] (mlk) at (6, -1.5) {};

  \node[call] (next0) at (3, 0) {\code{next}};

  \node[call] (call0) at (3.5, -1) {\code{call}};

  \draw     (mak)   -- (join1);
  \draw[->] (join1) -- (itk);
  \draw[->] (join1) -- (fnk);
  \draw     (itk1)  -- (join2);
  \draw     (fnk1)  -- (join2);
  \draw[->] (join2) -- (mak1);

  \draw[->] (itk)   -- (next0);
  \draw[->] (next0) -- (itk1); \draw[->] (next0) -- (vlk);

  \draw[->] (fnk)   -- (call0); \draw[->] (vlk) -- (call0);
  \draw[->] (call0) -- (fnk1); \draw[->] (call0) -- (mlk);

  \draw (1, .7) rectangle (5.5, -2.3);
  \node at (3.25, -2.5) {\code{Map::next}};
\end{tikzpicture}
\begin{tikzpicture}[scale=0.77, every node/.style={transform shape}]
  \node[state,label={$P_k$}]     (mak)  at (0.5,  -.5) {};
  \node[state,label={$P_{k+1}$}] (mak1) at (10, -.5) {};

  \coordinate (join1) at (1.25, -.5) {};
  \coordinate (join2) at (9.25, -.5) {};

  \node[state,label={$I_p$}]     (itp)  at (2,   0) {};
  \node[state,label={$I_{p+1}$}] (itp1) at (4,   0) {};
  \node                          (itpt) at (6,   0) {\dots};
  \node[state,label={$I_{p+x}$}] (itpn) at (8.5, 0) {};

  \node[state,label={below:$F_p$}]     (fnp)  at (2,   -1) {};
  \node[state,label={below:$F_{p+1}$}] (fnp1) at (4.5, -1) {};
  \node                                (fnpt) at (6.5, -1) {\dots};
  \node[state,label={below:$F_{p+x}$}] (fnpn) at (8.5, -1) {};

  \node[state,label={180:$v_p$}]     (vlp)  at (3.25, -.5) {};
  \node[state,label={180:$v_{p+1}$}] (vlp1) at (5.25, -.5) {};
  \node[state,label={180:$v_{p+x}$}] (vlpn) at (7.25, -.5) {};

  \node[anchor=west] (fres0) at (3.5, -1.9) {\code{false}};
  \node[anchor=west] (fres1) at (5.5, -1.9) {\code{false}};
  \node[anchor=west] (fresn) at (7.5, -1.9) {\code{true}};

  \node[state,label={$m_k$}] (mlk) at (10, -1.2) {};

  \node[call] (next0) at (3, 0) {\code{next}};
  \node[call] (next1) at (5, 0) {\code{next}};
  \node[call] (nextn) at (7, 0) {\code{next}};

  \node[call] (call0) at (3.5, -1) {\code{call}};
  \node[call] (call1) at (5.5, -1) {\code{call}};
  \node[call] (calln) at (7.5, -1) {\code{call}};

  \draw     (mak)   -- (join1);
  \draw[->] (join1) -- (itp);
  \draw[->] (join1) -- (fnp);
  \draw     (itpn)  -- (join2);
  \draw     (fnpn)  -- (join2);
  \draw[->] (join2) -- (mak1);

  \draw[->] (itp)   -- (next0);
  \draw[->] (next0) -- (itp1); \draw[->] (next0) -- (vlp);
  \draw[->] (itp1)  -- (next1);
  \draw[->] (next1) -- (itpt); \draw[->] (next1) -- (vlp1);
  \draw[->] (itpt)  -- (nextn);
  \draw[->] (nextn) -- (itpn); \draw[->] (nextn) -- (vlpn);

  \draw[->] (fnp)   -- (call0); \draw[->] (vlp)   -- (call0);
  \draw[->] (call0) -- (fnp1);  \draw[->] (call0) -- (fres0);
  \draw[->] (fnp1)  -- (call1); \draw[->] (vlp1)  -- (call1);
  \draw[->] (call1) -- (fnpt);  \draw[->] (call1) -- (fres1);
  \draw[->] (fnpt)  -- (calln); \draw[->] (vlpn)  -- (calln);
  \draw[->] (calln) -- (fnpn);  \draw[->] (calln) -- (fresn);

  \draw[->] (vlpn) -- (mlk);

  \draw (1, .7) rectangle (9.5, -2.3);
  \node at (5.25, -2.5) {\code{Filter::next}};
\end{tikzpicture}
\end{adjustbox}
\caption{State diagrams for a call to \code{Map::next} (left) and a call to \code{Filter::next} (right).}
\label{fig:map-filter}
\end{figure}

%% file: figures/dia78.tex
\begin{center}\begin{tikzpicture}
  \node                          (itt1) at (1, 0) {\dots};
  \node[state,label={$I_k$}]     (itk)  at (2, 0) {};
  \node[state,label={$I_{k+1}$}] (itk1) at (4.5, 0) {};
  \node                          (itt2) at (7, 0) {\dots};
  \node[state,label={$I_l$}]     (itl)  at (8, 0) {};
  \node                          (itt3) at (11, 0) {\dots};

  \node[state,label={$v_k$}] (vlk) at (4.5, -.8) {};
  \node                      (vlt) at (7.5, -.8) {\dots};
  \node                      (vlt2) at (10.5, -.8) {\dots};

  \node[call] (next0) at (3, 0) {\code{next}};
  \node[call] (next1) at (6, 0) {\code{next}};
  \node[call] (next2) at (9, 0) {\code{next}};

  \draw[->] (itt1)  -- (itk);
  \draw[->] (itk)   -- (next0);
  \draw[->] (next0) -- (itk1); \draw[->] (next0.345) -- (vlk);
  \draw[->] (itk1)  -- (next1);
  \draw[->] (next1) -- (itt2); \draw[->] (next1.345) -- (vlt);
  \draw[->] (itt2)  -- (itl);
  \draw[->] (itl)   -- (next2);
  \draw[->] (next2) -- (itt3); \draw[->] (next2.345) -- (vlt2);

  \node[anchor=west] (step1) at (1.55, -1.3) {\phantom{\code{step(}}};
  \node[anchor=west] (step2) at (1.55, -1.3) {\phantom{\code{step($I_k$, }}};
  \node[anchor=west] (step3) at (1.55, -1.3) {\phantom{\code{step($I_k$, $I_{k+1}$, }}};
  \node[anchor=west] (step)  at (1.5, -1.5)  {\code{step($I_k$, $I_{k+1}$, $v_k$)}};

  \draw[dotted,thick] (itk.south)  -- (step1.east);
  \draw[dotted,thick] (itk1.south) -- (step2.east);
  \draw[dotted,thick] (vlk.south)  -- (step3.east);

  \node[anchor=west] (lead1) at (5.85, -1.3) {\phantom{\code{leadsto(}}};
  \node[anchor=west] (lead2) at (5.85, -1.3) {\phantom{\code{leadsto($I_{k + 1}$, }}};
  \node[anchor=west] (lead)  at (5.8, -1.5)  {\code{leadsto($I_{k + 1}$, $I_l$)}};

  \draw[dotted,thick] (itk1.south) -- (lead1.east);
  \draw[dotted,thick] (itl.south) -- (lead2.east);
\end{tikzpicture}\end{center}

%% file: sections/iterators.tex
In this section, we define a concrete Rust-like specification language for expressing the specifications required by our methodology ($\rightarrow$ \secref{iterators-syntax}) and describe the idea of a \emph{snapshot abstraction} to express properties of the entire state of a data structure ($\rightarrow$ \secref{iterators-snapshots}). We use the language to instantiate the methodology introduced in~\secref{methodology} ($\rightarrow$ \secref{iterators-specext}), and finally illustrate the methodology on a simple, unchained iterator ($\rightarrow$ \secref{iterators-simple}).

\subsection{Specification language}\label{sec:iterators-syntax}

Contracts are attached to methods using Rust attributes for each kind of specification, \eg{} pre- and postconditions are annotated as follows:

\begin{rust}
#[requires(a ==> b)]   // method precondition
#[ensures(result > 0)] // method postcondition
fn example(a: bool, b: bool) -> i32 { ... }
\end{rust}

The expressions in parentheses after the \code{requires} or \code{ensures} keyword are side-effect-free Boolean Rust expressions, with a few (but powerful) extensions:

\begin{itemize}
  \item Implications, represented as \code{antecedent ==> consequent}.
  \item Universal and existential quantifiers, represented as:
  \begin{itemize}
    \item \code{forall(|qvar_ident: QvarType| body_of_quantifier)}
    \item \code{exists(|qvar_ident: QvarType| body_of_quantifier)}
  \end{itemize}
  \item Mathematical types, particularly \code{GhostSeq<T>}, a generic type representing a sequence of instances of type \code{T}.
  \item Logical equality, represented as \code{lhs === rhs}. This operation is distinct from the Rust equality \code{lhs == rhs} (note double equals in Rust) and represents a deep structural comparison, \ie{} a comparison of all the memory reachable from each operand. Further discussion can be found in~\secref{iterators-snapshots}.
  \item Call descriptions, represented as \code{F $\;\rightsquigarrow$ |args...| \{ pre \} \{ post \}}. This expression states that a call to the method \code{F} \emph{has happened}, the assertion \code{pre} held in the prestate of that call, the assertion \code{post} held in the poststate, and the arguments of the call are bound to the names \code{args...} and the return value to a reserved name \code{result} in the assertions \code{pre} and \code{post}. The specific case of a closure call is a call to the method \code{FnMut::call_mut}, with the closure state constrained using the pre- and poststate assertions. Similarly, the use of an iterator is a call to the method \code{Iterator::next}. This notation is based on prior work on closure verification~\cite{Wolff_Closures_2021}.
\end{itemize}

Ghost functions are functions used only from within specifications and are also allowed to use the extended expression syntax:

\begin{rust}
fn all_false(x: GhostSeq<bool>) -> bool {
    forall(|idx: usize| 0 <= idx && idx < x.len() ==> !x[idx])
}
\end{rust}

\subsection{Abstract data structure states}\label{sec:iterators-snapshots}

Specifications in our methodology concern properties of the \emph{state} of a data structure such as an iterator, a closure, or the values returned by an iterator. For instance, \code{produced} needs to contain the values returned during an iteration. Moreover, \code{step} and \code{leadsto} typically compare states of data structures, and there are examples that need to quantify over the states of data structures, for example, to express the effects of multiple closure calls. In all of these cases, the meanings of our specifications need to refer to the entire current state of a data structure in memory (such as an iterator), not just its identity.

To enable such specifications, we introduce a canonical abstraction of each data structure that captures its structure and values, but abstracts away concrete addresses. We refer to the abstract value of a data structure as its \emph{snapshot}. Since data structures in (safe) Rust are always tree-shaped, a snapshot can be represented easily as terms of a (mathematical) abstract data type. For instance, the snapshot of a \code{Point} object (\secref{bg-rust}) is a mathematical tuple of two integers.

It is common for separation logic specifications to relate the memory reachable from a given program expression to a user-defined abstract value using (possibly recursive) predicate definitions~\cite{Parkinson_SLAbstraction_2005}. Our snapshots are similar to these abstract values, but the abstraction is derived automatically from the declarations of types and fields.

Snapshots allow our specifications to conveniently express properties of entire states of data structures. For instance, logical equality (\code{===}) is simply a comparison of two snapshots. Similarly, function \code{produced} yields a snapshot that includes the snapshots of all values returned by the iterator so far. Snapshots also allow cleanly integrating state-dependent properties with logical features such as quantifiers and mathematical functions. For example, in Prusti~\cite{Prusti_2019}, following Smans et al.~\cite{Smans_Snapshots_2010}, ghost functions which operate on Rust data are transparently encoded to mathematical functions which operate on snapshots. The same approach has been applied to encoding quantifiers and call descriptions~\cite{Wolff_Closures_2021}, as used in our work (instead of quantifying over expressions in different states, one quantifies over their snapshots as mathematical values).

A (possibly surprising) consequence of the snapshot interpretation of our specifications is that a ghost function can take parameters that represent multiple versions of the \emph{same} data structure, which is not directly possible in actual Rust code. This allows us to encode two-state predicates such as \code{step} and \code{leadsto} as standard ghost functions that take two snapshots as arguments, representing the iterator state \emph{at different points} in the program execution.

Since snapshots are simply terms of a mathematical data type, they have a straightforward encoding into SMT, which partly addresses the automation challenge mentioned in the introduction. In fact, several automated program verifiers employ similar notions of snapshots internally in their encodings, for instance, Smallfoot~\cite{Smans_Snapshots_2010}, VeriFast~\cite{Jacobs_VeriFast_2011}, and Viper~\cite{Viper_2016}.

\subsection{Defining the specification components}\label{sec:iterators-specext}

In Rust, iterators are types which implement the \code{Iterator} trait. This is achieved with an explicit syntactic declaration (\code{impl Iterator for SomeType}) which also declares the element type and provides the definition of one required method: \code{next}. The implementation of this method can modify the current state of the iterator and must output an \code{Option} which may contain the next element, or it may be \code{None} to signal completion. The \code{Iterator} trait defines a large number of other methods, such as \code{map} or \code{filter}; these all have default implementations, which are inherited into any concrete iterator.

Conceptually, we can represent the four key components of our iterator specifications (see \secref{methodology-components}) as ghost functions of the \code{Iterator} trait. Because we provide no default implementations, every concrete iterator would then need to provide definitions for these functions. Our implementation uses an alternative approach that does not require any changes to functions declared in the \code{Iterator} trait, see \secref{impl}.

\begin{rust}
trait Iterator {
    fn produced(&self) -> GhostSeq<Self::Item>;
    fn step(i_p: &Self, i_c: &Self, r: &Option<Self::Item>) -> bool;
    fn leadsto(i_p: &Self, i_c: &Self) -> bool;
    fn completed(&self) -> bool;
    // other Iterator methods
}
\end{rust}

The \code{produced} method is a getter\footnote{Traits in Rust cannot declare fields or properties, thus we must use a getter method.} for a ghost sequence of values. The two-state predicates \code{step} and \code{leadsto} takes references to two copies (a previous version and a current version) of the iterator type. \code{step} additionally takes a reference to the returned value, represented in Rust as an \code{Option} to allow for \code{None} to signal the end of iteration.
As noted in~\secref{iterators-snapshots}, it is possible to give the two-state predicates \code{step} and \code{leadsto} different versions of the same instance using snapshots.
Finally, \code{completed} defines the stopping condition. Since these methods are ghost code, we allow them to use non-executable constructs such as quantifiers in their definitions in implementations of the \code{Iterator} trait.

Finally, we express the proof obligations from \secref{methodology} as postconditions of the \code{next} method, which must be satisfied by each implementation of \code{next}. Here, \code{&result} denotes a shared reference to the value returned by the method.

\begin{rust}
trait Iterator {
(*\codepoint{$Q_1$}*) #[ensures(IteratorSpec::step(old(&self), &self, &result))]
(*\codepoint{$Q_2$}*) #[ensures(old(self.completed()) == (result === None))]
(*\codepoint{$Q_3$}*) #[ensures(old(self.completed()) ==> (
        self.produced() === old(self.produced())
    ))]
(*\codepoint{$Q_4$}*) #[ensures(!old(self.completed()) ==> (
           result === Some(self.produced().last())
        && self.produced().len() == old(self.produced().len()) + 1
        && GhostSeq::prefix(old(self.produced()), self.produced())
    ))]
(*\codepoint{$Q_5$}*) #[ensures(old(self.completed()) ==> self.completed())]
    fn next(&mut self) -> Option<Self::Item>;
    // ...
}
\end{rust}

This specification uses logical equality \code{===}, described in~\secref{iterators-syntax}. Using \code{==} would require a \code{PartialEq} trait-bound on the iterator's elements, restricting its interface (not all iterators' elements need have a Rust definition of equality).

\subsection{Example} \label{sec:iterators-simple}

We illustrate our methodology by specifying a simple iterator returning a range of consecutive numbers. In Rust, we can define such an iterator like this:

\begin{rust}
struct Counter { (*\codepoint{A}*)
    pos: i32,
    end: i32,
}
impl Counter {
    #[ensures(Iterator::leadsto(&result, &result))]
    fn new(end: i32) -> Self { ... } (*\codepoint{B}*)
}
impl Iterator for Counter { (*\codepoint{C}*)
    type Item = i32; (*\codepoint{D}*)
    fn next(&mut self) -> Option<Self::Item> { (*\codepoint{E}*)
        if self.pos <= self.end {
            self.pos += 1;
            Some(self.pos - 1)
        } else {
            None
        }
    }
    // ...
}
\end{rust}

The code above first \codepoint{A} defines a type which represents our iterator. Its state contains two variables: its current position and its stopping point. \codepoint{B} defines a convenience constructor for the counter, which must ensure that the two-state invariant \code{leadsto} is (reflexively) satisfied to begin with. \codepoint{C} is the declaration that marks \code{Counter} to be an iterator, consisting of a declaration of the type of elements emitted by this iterator \codepoint{D}, and an implementation of the \code{next} method \codepoint{E}. The latter checks if the limit has been reached yet: if so, no more items are emitted (as signalled by a \code{None} return value), otherwise the internal position is updated and its old value is returned (wrapped in a \code{Some} to distinguish it from the previous case).

To specify this iterator, we provide definitions of the four ghost methods declared in the \code{Iterator} trait, starting with \code{completed}:

\begin{rust}
    fn completed(&self) -> bool {
        !(self.pos <= self.end)
    }
\end{rust}

This definition is sufficient to show that a counter with a non-empty range will return at least one value:

\begin{rust}
let mut counter = Counter::new(2);
assert!(counter.next().is_some());
\end{rust}

The two-state postcondition \code{step} defines how the state is updated. Here, \code{i_p} and \code{i_p} refer to the iterator snapshot before and after the execution of \code{next}:

\begin{rust}
    fn step(i_p: &Self, i_c: &Self, r: &Option<Self::Item>) -> bool {
        !i_p.completed() ==> (
               i_c.pos == i_p.pos + 1
            && *res === Some(i_p.pos) (*\codepoint{A}*)
        )
        && i_p.end == i_c.end (*\codepoint{B}*)
    }
\end{rust}

This definition is sufficient to prove assertions in a non-looping client:

\begin{rust}
let mut counter = Counter::new(1);
assert!(counter.next().unwrap() == 0);
assert!(counter.next().unwrap() == 1);
assert!(counter.next().is_none());
\end{rust}

Note that the conjunct \codepoint{A} would follow from knowledge about the \code{produced} sequence, and the conjuncts \codepoint{B} are actually stating an invariant that is true between any two states of the counter, not just consecutive ones. We can thus simplify the definition of \code{step} and provide a definition of \code{leadsto}:

\begin{rust}
    fn step(i_p: &Self, i_c: &Self, r: &Option<Self::Item>) -> bool {
        !i_p.completed() ==> (i_c.pos == i_p.pos + 1)
    }
    fn leadsto(i_p: &Self, i_c: &Self) -> bool {
    (*\codepoint{C}*)    i_p.end == i_c.end
    (*\codepoint{D}*) && i_p.pos <= i_c.pos
    (*\codepoint{E}*) && 0 <= i_c.pos && i_c.pos <= i_c.end + 1
    (*\codepoint{F}*) && i_c.produced().len() == i_c.pos as usize
    (*\codepoint{G}*) && forall(|x: i32| 0 <= x && x < i_c.pos ==>
            i_c.produced()[x as usize] == x)
    }
\end{rust}

In this definition, \codepoint{C} \code{end} remains constant. \codepoint{D} \code{pos} is monotonically increasing, and \codepoint{E} it remains within bounds. \codepoint{F} the number of produced elements is given by the difference between the current position and the lower bound. Finally, \codepoint{G} the value of every produced element can be defined by its position in the sequence.

The advantage of this more verbose definition is that \code{Counter} can be used across unboundedly many calls, such as in a loop:

\begin{rust}
let mut counter = Counter::new(90);
let val_pre = counter.next().unwrap();
assume!(n < 80);
for i in 0..n {
    counter.next().unwrap();
}
assert!(counter.next().unwrap() > val_pre);
\end{rust}

At this point we have specified a simple iterator. We have used all four of the methodology components we defined in~\secref{methodology}. Although the approach is relatively heavy for \code{Counter} (the definition of \code{leadsto} in particular), we will shortly see it pays off when considering more advanced (and idiomatic) cases. A similar specification can be used for iterating over a slice or \code{Vec}.

%% file: sections/chains.tex
In this section, we gradually introduce our technical solutions to the second challenge described in \secref{introduction}: modular reasoning about iterator chains. To introduce iterator adapters in Rust, we specify a simple doubling iterator adapter ($\rightarrow$ \secref{chains-adapt-one}). Approaching the real \code{Map} type from the standard library, we specify an adapter with side-effectful closures ($\rightarrow$ \secref{chains-side}). For the real \code{Map} specification, we also need to account for transitive side effects, \ie{} side effects of any nested iterators ($\rightarrow$ \secref{chains-adapt-many}).

\subsection{Simple iterator adapters}\label{sec:chains-adapt-one}

In Rust, iterator adapters are types which wrap an instance of the \code{Iterator} trait while implementing the \code{Iterator} trait themselves. When the adapter's \code{next} method is invoked, it calls the previous iterator's \code{next} method some number of times, adapting its result. As an example, the \code{Map} adapter applies a closure to every element produced by the previous iterator, while the \code{Filter} adapter uses a closure as a logical predicate to decide whether each value from the previous iterator should be returned or not.

The Rust standard library provides many such iterator adapters for a variety of common use cases. The iterator chain in the code example shown in~\secref{introduction} is actually composed of iterator adapters, albeit hidden behind some convenience syntax (\eg{} \code{x.map(...)} wraps the iterator \code{x} in a \code{Map} adapter).

As stated before, our goal is to write single, generic specifications, that apply regardless of the exact iterator that precedes the adapter in the chain or any specific closure parameters. We first show that we can specify simple adapters by characterising their \code{produced} sequence in terms of the \code{produced} sequence of the previous iterator. For example, consider the following \code{Double} iterator adapter which simply doubles the values coming from the previous iterator:

\begin{rust}
struct Double<I (*\codepoint{A}*)> {
    prev_iter: I, (*\codepoint{B}*)
}
impl<I: Iterator<Item = i32>> Iterator for Double<I> {
    type Item = i32;
    fn next(&mut self) -> Option<Self::Item> {
        self.prev_iter.next().map(|v| v * 2) (*\codepoint{C}*)
    }
}
\end{rust}

This iterator adapter is able to adapt any iterator type that yields integers. It thus has a type parameter \codepoint{A}, which stands for the type of the previous iterator. To be able to refer and mutate the previous iterator, the struct contains (or \emph{owns}) the previous iterator \codepoint{B}. Finally, we implement the \code{Iterator} trait for \code{Double} with a straightforward mapping\footnote{The \code{map} method on an \code{Option} type applies the given closure on the value contained in a \code{Some}, but leaves \code{None} intact.} of values \codepoint{C}.

We omit the straightforward definition of \code{step}. A possible (abridged) definition of \code{leadsto} and \code{completed} is:

\begin{rust}
fn leadsto(i_p: &Self, i_c: &Self) -> bool {
(*\codepoint{A}*)    i_c.produced().len() == i_c.prev_iter.produced().len()
(*\codepoint{B}*) && forall(|idx: usize| idx < i_c.produced().len() ==>
        i_c.produced()[idx] == i_c.prev_iter.produced()[idx] * 2)
}
fn completed(&self) -> bool {
(*\codepoint{C}*) self.prev_iter.completed()
}
\end{rust}

In the definition of \code{leadsto}, we \codepoint{A} state that the length of the \code{produced} sequence of the previous iterator and the adapter \code{Double} is the same, and we \codepoint{B} use a quantifier to relate each element. We also \codepoint{C} re-use the previous iterator's stopping criterion. This definition is sufficient to verify properties of the values produced when \code{Double} is used over an iterator (such as the previously defined \code{Counter}):
\begin{rust}
let counter = Counter::new(2);
let mut double_iter = Double::new(counter);
assert!(double_iter.next().unwrap() == 0);
assert!(double_iter.next().unwrap() == 2);
assert!(double_iter.next().unwrap() == 4);
assert!(double_iter.next().is_none());
\end{rust}
The contract of \code{next} implies that if a \code{Some} result is obtained, \code{produced} was extended by one element. For each of the first three calls to \code{next} the verifier thus learns about a new element of this sequence, the value of which is related to the corresponding element of the previous iterator's \code{produced} sequence.
This establishes a concrete relationship between the elements yielded by \code{Counter} and those yielded by \code{Double}, sufficient to imply all asserted properties about these resulting values.

\subsection{Side-effectful closures in adapters}\label{sec:chains-side}

Although \code{Double} is indeed an iterator adapter, it is rather simple. The action it performs is functionally a fixed one-to-one \emph{mapping} of values. The \code{Map} iterator adapter performs exactly this action, but let's the user supply the function to be applied. To this end, the \code{Map} type has two type parameters: one to represent the previous iterator in the chain and one to represent the closure\footnote{The actual declaration has a third type parameter to represent the return type of the closure, omitted here for brevity.}:

\begin{rust}
struct Map<I, F> {
    prev_iter: I, // the wrapped iterator
    f: F,         // the closure parameter
}
\end{rust}

When mapping values from the previous iterator, \code{Map} invokes the closure:

\begin{rust}
fn next(&mut self) -> Option<Self::Item> {
    self.prev_iter.next().map(&mut self.f)
}
\end{rust}

To provide a generic specification of \code{Map}, we must account for the side effects of the closure call, even when the exact type of the closure is unknown; this property of side effects on the \emph{closure}'s mutable state cannot be captured using the \code{produced} sequences alone. Instead, we use a call description to connect the effect of an iteration step to the effects of a call to the closure itself:

\begin{rust}
fn step(i_p: &Self, i_c: &Self, r: &Option<Self::Item>) -> bool {
(*\codepoint{A}*) !i_p.prev_iter.completed() ==>
    (*\codepoint{B}*) FnMut::call_mut (*$\rightsquigarrow$*) |cl_self, arg|
        (*\codepoint{C}*) {    i_p.f === cl_self
        (*\codepoint{D}*)   && arg === i_c.prev_iter.produced().last() }
        (*\codepoint{E}*) {    i_c.f === cl_self
        (*\codepoint{F}*)   && r === Some(result) })
}
\end{rust}
For each step we know (if the previous iterator has not completed yet\codepoint{A}) that \codepoint{B} there is a call to the closure, with \codepoint{C} the original state of the closure stored in field \code{f} of the original state of \code{Map}, and \codepoint{E} the new state of the closure stored in field \code{f} of the new state of \code{Map}. For the call, \codepoint{D} the argument given to the closure is the last element yielded by the previous iterator, and \codepoint{F} the element yielded by \code{Map} (\code{r}) is the result of the closure (\code{Some(result)}).

This definition of \code{step} allows us to verify the \emph{side effects} of individual calls to \code{next} on the corresponding closure's \emph{captured state}:

\begin{rust}
let counter = Counter::new(2);
let mut sum = 0;
let cl = #[ensures(sum == old(sum) + arg && result == arg * 2)]
    |arg| { sum += arg; arg * 2 };
let mut map_iter = Map::new(counter, cl);
assert!(map_iter.next().unwrap() == 0);
assert!(map_iter.next().unwrap() == 2);
assert!(sum == 1);
\end{rust}

The final assertion (after \code{map_iter} and \code{cl} expire) follows from the call description describing the call \code{Map} makes to the closure, which establishes the postcondition of the concrete closure known concretely in this client code\footnote{We assume the closures themselves to be annotated with standard pre/postcondition specifications; for simple closures such as this one, the required specifications could be inferred by a static analysis, but this is orthogonal to our presented technique.}.

\subsection{Multiple adapters}\label{sec:chains-adapt-many}

The \code{Map} specification in the previous subsection refers to the values produced by the previous iterator, but does not account for changes to its state. To address \emph{transitive} side effects of \code{next} calls (\ie{} those on closures earlier in the iterator chain), we must directly capture information about the calls to the previous iterator. Our call description feature can also be used to describe this necessary connection, this time relating each change to the current iterator to the corresponding pre- and poststate of a call to \code{next} on the \emph{previous} iterator, all the while keeping the specification generic with respect to the previous iterator's type. The following definition of \code{step} exemplifies this powerful idiom, using two nested call descriptions:

\begin{rust}
fn step(i_p: &Self, i_c: &Self, map_res: &Option<Self::Item>) -> bool {
(*\codepoint{A}*) Iterator::next (*$\rightsquigarrow$*) |it_self|
    (*\codepoint{B}*) { i_p.prev_iter === it_self }
    (*\codepoint{C}*) { let prev_iter_res = result;
    (*\codepoint{D}*)      i_c.prev_iter === it_self
    (*\codepoint{E}*)   && (prev_iter_res.is_some() ==>
        (*\codepoint{F}*) FnMut::call_mut (*$\rightsquigarrow$*) |cl_self, arg|
            {    i_p.f === cl_self
              && Some(arg) === prev_iter_res }
            {    i_c.f === cl_self
              && map_res === Some(result) } )
    (*\codepoint{G}*)   && prev_iter_res.is_none() ==> map_res.is_none() }
}
\end{rust}

In this version of \code{step}, we begin by \codepoint{A} describing that a call to the previous iterator \emph{will happen}, where (\codepoint{B}) the original state of the previous iterator is stored in field \code{prev_iter} of the original state of the \code{Map} iterator, and \codepoint{D} the new state of the previous iterator is as stored in field \code{prev_iter} of the resulting state of \code{Map}. To avoid confusion, we \codepoint{C} used a let expression to name the result of the previous iterator \code{prev_iter_res}; the result returned from the \code{Map} is \code{map_res}. \codepoint{E} If the previous iterator returned an element, we \codepoint{F} describe how this relates a call to the closure, as before. \codepoint{G} Otherwise, no element is returned from the \code{Map} iterator either.

The overall structure of this definition of \code{step} mirrors the \code{Map} model presented in~\figref{map-filter} (left): each call description corresponds to one box in the diagram, and values from the previous iterator flow through the closure.

With the richer specifications, we can prove facts about transitive side effects:

\begin{rust}
let counter = Counter::new(2);
let mut sum = 0;
let cl_1 = #[ensures(sum == old(sum) + arg && result == arg + 1)]
    |arg| { sum += arg; arg * 2 };
let map_iter_1 = Map::new(counter, cl_1);
let cl_2 = #[ensures(result == arg * 2)] |arg| arg * 2;
let mut map_iter_2 = Map::new(map_iter_1, cl_2);
assert!(map_iter_2.next().unwrap() == 2);
assert!(sum == 0);
\end{rust}

The final assertion requires that the knowledge about the effect of \code{cl_1} on its captured variable is propagated through the invocation of \code{map_iter_1}, which happens inside \code{map_iter_2}. Note that the specification is the same for both \code{Map} iterators, and that \code{map_iter_2} in particular cannot even know that the previous iterator in the chain is related to a closure.

Finally, reasoning about \emph{unboundedly many iteration steps} (\eg{} when calling \code{next} in a loop), simply requires us to suitably define \code{leadsto} for \code{Map}. To store the intermediate states of the closures and the iterators, we add a ghost sequence field \code{st: GhostSeq<(F, I)>} to \code{Map}, where every element is a tuple containing a closure state and an iterator state.

\begin{rust}
fn leadsto(i_p: &Self, i_c: &Self) -> bool {
       i_c.produced().len() == i_c.prev_iter.produced().len()
(*\codepoint{A}*) && i_p.st.is_prefix_of(i_c.st)
(*\codepoint{B}*) && i_c.st.len() == i_c.produced().len() + 1
(*\codepoint{C}*) && i_c.st.last() === (i_c.f, i_c.prev_iter)
    && forall(|idx: usize| idx < i_c.produced().len() ==>
    (*\codepoint{D}*) I::next (*$\rightsquigarrow$*) |iter_self|
            {    i_c.st[idx].1 === iter_self }
            {    i_c.st[idx + 1].1 === iter_self
              && result === Some(i_c.prev_iter.produced()[idx])
              && FnMut::call_mut (*$\rightsquigarrow$*) |cl_self, arg|
                     {    i_c.st[idx].0 === cl_self
                       && arg === i_c.prev_iter.produced()[idx] }
                     {    i_c.st[idx + 1].0 === cl_self
                       && result === i_c.produced()[idx] }})
}
\end{rust}

In this definition of \code{leadsto}, we establish a connection between the ghost sequence \code{st} and the concrete data stored in the \code{Map} struct. In particular, \codepoint{B} states there are as many intermediate states as there are yielded elements, plus one for the current state. \codepoint{C} The last tuple in the \code{st} sequence corresponds to the current data stored in \code{Map}. For each yielded element \codepoint{D} we re-use our nested call descriptions to establish the connection between consecutive intermediate states. We also \codepoint{A} state that the state sequence is expanded monotonically, \ie{} the intermediate states in any previous version of \code{Map} are a prefix of the intermediate states of any newer version of \code{Map}.

Equipped with such a specification, we can prove code which uses a \code{Map} parameterised by a closure with captured state, applied to an unbounded number of elements, such as this vector summation:

\begin{rust}
#[ensures(result == GhostSeq::of_vec(&vec).sum())]
fn sum_vec(vec: Vec<i32>) -> i32 {
    let vec_vals = GhostSeq::of_vec(&vec);
    // ghost state of closure
    let mut pos = 0usize;
    let mut sum = 0;

    // summation closure
    let cl = #[requires(pos < vec_vals.len() && x == vec_vals[pos])]
             #[ensures(pos == old(pos) + 1 && ret == sum)]
             #[invariant(   0 <= pos && pos <= vec_vals.len()
                         && sum == vec_vals[0..pos].sum())]
             |x| { sum += x; pos += 1; sum };

    // iterate values of vector
    let mut map_iter = vec.into_iter().map(cl);
    #[invariant(pos == map.prev_iter.pos)]
    for el in map_iter {}

    sum
}
\end{rust}

In this example, \code{leadsto} is maintained for \code{map_iter} throughout the loop, which also implies the closure history invariant. Once the loop exits, it is known that the position of the vector iterator reached the end of the vector, and so \code{sum == vec_vals[0..pos].sum()} collapses to \code{sum == vec_vals.sum()}.

%% file: sections/impl.tex
So far, we assumed that Rust traits, including those from the standard library, can be extended with additional methods. To avoid such changes to existing code, Spec\# proposed \emph{out-of-band contracts} as a mechanism to attach specifications to libraries. In this section, we present a refined notion of out-of-band contracts  that allow one to attach contracts \emph{conditionally}, depending on whether a type implements a trait.

\subsection{Specification extension traits}\label{sec:impl-specext}

Instead of changing Rust's \code{Iterator} trait, we declare the four (specification-only) methods of our methodology in a new trait:

\begin{rust}
trait IteratorSpec : Iterator { (*\codepoint{A}*)
    fn produced(&self) -> GhostSeq<Self::Item (*\codepoint{B}*)>;
    fn leadsto(i_p: &Self, i_c: &Self) -> bool;
    fn step(i_p: &Self, i_c: &Self,
        r: &Option<Self::Item (*\codepoint{C}*)>) -> bool;
    fn completed(&self) -> bool;
}
\end{rust}

In the above, we \codepoint{A} declare \code{IteratorSpec} and use \code{Iterator} as its super-trait. This means that any type that implements \code{IteratorSpec} must also implement \code{Iterator}. Apart from declaring the same methods we have already seen before, one notable feature is \codepoint{B}, \codepoint{C} re-using the associated type \code{Item} from the \code{Iterator} supertrait, such that the implementation of \code{IteratorSpec} does not need to repeat the element type that is already declared in \code{Iterator}.

Declaring methodology components in traits should be familiar to Rust programmers because they represent a shared behaviour. Like regular traits, any given implementation of \code{IteratorSpec} can define these methods to specify concrete iterators.

In our implementation, the method declarations in \code{IteratorSpec} are additionally annotated with either \code{#[predicate]} or \code{#[pure]}. Both annotations indicate that the given method is usable in specifications, and \code{#[predicate]} additionally allows the use of non-executable features such as quantifiers.

\subsection{Type-dependent contracts}\label{sec:impl-typedep}

The specification extension trait \code{IteratorSpec} allows us to remove the four function declarations from the \code{Iterator} trait, but does not provide a way to express the postconditions of the \code{next} method (Rust does not allow a sub-trait to override methods of a super-trait). To provide a specification for \code{next} without changing the \code{Iterator} trait, we use an out-of-band contract for the trait, that is, a specification provided in a separate file. For example, we could (erroneously) specify \code{Iterator::next} to always return a result:

\begin{rust}
#[extern_spec] // out-of-band contract
trait Iterator {
    #[ensures(result.is_some())]
    fn next(&mut self) -> Option<Self::Item>;
}
\end{rust}

However, such an extern spec for \code{Iterator} cannot use the methods from \code{IteratorSpec}, because not every implementation of \code{Iterator} also implements \code{IteratorSpec}. To solve this issue, we introduce type-dependent contracts, that is, contracts that apply \emph{only} to implementations of a given trait. The following extern spec for \code{Iterator} uses this feature to impose postconditions on \code{next} \emph{only} for implementations that also implement \code{IteratorSpec}:

\begin{rust}
#[extern_spec]
trait Iterator { (*\codepoint{A}*)
    #[type_dependent(Self: IteratorSpec, [ (*\codepoint{B}*)
    (*\codepoint{$Q_1$}*) ensures(IteratorSpec::step(old(&self), &self, &result)),
    (*\codepoint{$Q_2$}*) ensures(old(self.completed() == (result === None))),
    (*\codepoint{$Q_3$}*) ensures(old(self.completed()) ==> (
            self.produced() === old(self.produced())
        )),
    (*\codepoint{$Q_4$}*) ensures(!old(self.completed()) ==> (
               result === Some(self.produced().last())
            && self.produced().len() == old(self.produced().len()) + 1
            && GhostSeq::prefix(old(self.produced()), self.produced())
        )),
    (*\codepoint{$Q_5$}*) ensures(old(self.completed()) ==> self.completed()),
    ])]
    fn next(&mut self) -> Option<Self::Item>;
}
\end{rust}

As before, \codepoint{A} an extern spec is added to the \code{Iterator} trait. The \codepoint{B} \code{type_dependent} attribute introduces postconditions \codepoint{$Q_1$}-\codepoint{$Q_5$} only for implementations that also implement \code{IteratorSpec}. As a result, it is allowed to refer to methods of \code{IteratorSpec}, \eg{} to call \code{self.completed()}.

Note that the type-dependent contract is defined on the type \emph{being extended}, \ie{} it is part of the specification of \code{Iterator::next}, not part of the specification extension trait \code{IteratorSpec}. This decision is motivated by Rust's \emph{coherence} rules.

To ensure that our type-dependent contracts are sound, we require that any type-dependent contract refines its base contract. This consists of the usual \emph{behavioural subtyping} checks: the precondition must be weakened and the postcondition must be strengthened.

%% file: sections/soundness.tex
Our methodology builds on standard techniques to reason about method implementations and calls, and on call descriptions to reason about closure calls~\cite{Wolff_Closures_2021}.  Consequently, its soundness relies on a sound verification technique for these features. Such techniques support two-state postconditions such as our \code{step} predicate (typically expressed via logical variables or old-expressions), which are checked for each method implementation and assumed for each call.

The only non-standard component of our methodology is the \code{leadsto} predicate, which may be assumed between any two iterator states occurring during an iteration. If \code{next} is the only method mutating the iterator state, this assumption is justified by the fact that each call to \code{next} satisfies \code{step}, and \code{leadsto} includes the reflexive, transitive closure of \code{step}. Consequently, \code{leadsto} holds between two iterator states, no matter how many calls to \code{next} occur between those states.

If the iterator state may be modified by methods other than \code{next}, one can interpret \code{leadsto} as a standard two-state invariant on the iterator type. Such an invariant must be maintained by all mutating methods, such that the above inductive argument applies.

%% file: sections/evaluation.tex
We implemented our technique as a prototype extension to the Prusti verifier~\cite{Prusti_2019}. Following the design laid out in \secref{iterators-syntax}, our iterator methodology is implemented primarily in user-facing Rust code (which can be packaged into a standard library of specifications) and not as an \emph{ad hoc} feature of Prusti.

To enable annotating key standard library types, we added support for declaring external specifications for trait types as well as for our novel type-dependent contracts. Most other extensions were routine; this suggests that layering our methodology onto existing tools is fairly lightweight. Due to this lightweight integration, we don't treat \code{leadsto} as a built-in type invariant, but rather as a postcondition on \code{next}; intended properties such as transitivity are therefore not checked by default (but proofs relying on them would of course fail otherwise).

\begin{table}[t]
  \begin{adjustbox}{center}
  \begin{tabular}{@{}lrrr@{}} \toprule
    Test (implementation) & LoS & LoC & VT (s) \\ \midrule
\code{counter} & 32 & 30 & 9.53 \\
\code{double} & 43 & 25 & 9.95 \\
\code{filter.vpr} & 90* & 109* & 18.24* \\
\code{map} & 67 & 38 & 42.12 \\
\code{option_intoiter} & 36 & 19 & 7.41 \\
\code{vec_intoiter} & 42 & 13 & 7.05 \\
\code{zip} & 79 & 32 & 84.46 \\
\bottomrule
  \end{tabular}
  \quad
  \begin{tabular}{@{}lrrr@{}} \toprule
    Test (client code) & LoS & LoC & VT (s) \\ \midrule
\code{counter} & 0 & 14 & 11.53 \\
\code{double} & 1 & 6 & 10.16 \\
\code{filter.vpr} & 10* & 27* & 5.68* \\
\code{map} & 14 & 22 & 79.78 \\
\code{option_intoiter} & 0 & 4 & 6.96 \\
\code{vec_intoiter} & 4 & 19 & 16.48 \\
\code{zip} & 2 & 6 & 67.12 \\
\bottomrule
  \end{tabular}
  \end{adjustbox}
  \vspace{5pt}
  \caption{\label{table:eval}Evaluation. \emph{LoS} and \emph{LoC} represent the line of specifications and lines of code, respectively. \emph{VT} represents the verification time, measured as the wall-clock runtime averaged over 7 runs using an Intel Core i9-10885H 2.40GHz CPU with 16~GiB of RAM, excluding the slowest and fastest runs. Test cases ending in \code{.vpr} were encoded manually into Viper: our methodology supports these test cases, but issues in the underlying Prusti tool (independent from the contributions of this paper) currently prevent our implementation from supporting the analogous examples in Rust. These Viper encoded examples are more verbose and run through a simpler tool chain; for these reasons we mark the data with *s in the tables here.}
\end{table}

We evaluated our work on a number of challenging test cases, modelling various combinations of idiomatic iterators found in the Rust standard library, as well as custom iterator implementations discussed in this paper. The results of our evaluation are shown in \tableref{eval} (in terms of lines of specification, code and verification times). Generally, the specification overhead is heavier (roughly one-to-one with code) for the generic library functions such as \code{Map}, but these specifications need only be written once. Importantly, for client code \emph{using} these iterators, the specification overhead is typically lighter. A substantial body (roughly 340 LoC) of common specifications were also necessary as our implementation neither builds in pre-defined support for common types such as \code{Option}, or our new \code{GhostSeq} type. These specifications need only be written once can could in principle be added as a ``standard library'' of specifications. We consider the verification times using our prototype implementation to be generally reasonable, but with some expensive outliers; we suspect that these require some additional effort to control quantifier instantiation in the underlying solver~\cite{AxiomProfiler_2019}.

%% file: sections/rw.tex
The two works most closely related to this paper (Pereira et al.~\cite{Pereira_Iterators_2018} and Wolff et al.~\cite{Wolff_Closures_2021}) were already discussed in detail in \secref{introduction}.

One of the first formulations of the challenges of specifying and verifying iterators appeared in the 2006 SAVCBS workshop~\cite{SAVCBSProceedings_2006} (although the problem has been studied for longer), which posed the research question in the context of Java/C\# collections and iterators over such collections. Various papers were submitted as solutions to the challenge: Jacobs~\cite{Jacobs_Iterators_2006} focuses on C\# iterators, which are suspendable functions that can ``yield'' values to a consumer. The iteration itself is characterised by invariants over the sequence of produced values, similarly to Pereira~\cite{Pereira_Iterators_2018}. C\# iterators are analogous to Rust generators (or ``coroutines'') which can be turned into an \code{Iterator} instance\footnote{Since generators and asynchronous code in Rust is still unstable in general, such adaptation is not yet part of the standard library. It is possible with relatively little boilerplate code, see \url{https://stackoverflow.com/questions/16421033}.}. Although Prusti does not yet support Rust generators, our methodology should apply to this case as well: the state of an iterator would become the state of the generator and the \code{leadsto} two-state type invariant would be maintained over the generator.

Krishnaswami~\cite{Krishnaswami_Iterators_2006} proposed specifications for Java-like iterators in a higher-order separation logic. The application of separation logic facilitates proofs of memory safety, and prevents concurrent modifications of a collection while an iterator over it exists. However, functional specifications are left as abstract higher-order predicates, which are problematic for automation.

Creusot~\cite{Denis_Creusot_2022} is another Rust verifier. Its prophetic approach to encoding Rust references simplifies the specification of some reborrowing patterns, which means the signature of traits such as \code{IterMut} are supported, unlike in Prusti at the time of writing. Creusot's public tests include functional specifications based on the Pereira methodology (both tools are based on Why3). If the contributions of this paper were adapted to Creusot, we believe it would be possible to specify richer properties of side-effectful iterator chains in this tool as well, as enabled by our novel methodology. More generally, our technique is largely agnostic as to the underlying verifier and its handling of \eg{} the program memory, provided that our snapshot interpretation of specifications can be appropriately supported.

%% file: sections/conclusion.tex
We have presented a novel methodology for modularly specifying and verifying the complex iterator patterns found in modern programming languages. This methodology is designed to be compatible with basic techniques for reasoning about side-effectful programs, such as Rust's ownership system and formal techniques such as separation logic; the underlying technical requirements are commonly found in many deductive verifiers. We have evaluated our methodology in Rust, which has rich iterator support in its standard library, as well as a type system which can be used to automatically take care of these ownership requirements. Applying our methodology in languages without such a type system would require specifications to govern side-effects, but the adaptation of our novel methodology would nonetheless be straightforward.

To ensure our methodology is usable in real-world codebases and integrates well with other verification efforts, we have prioritised modularity in our extension to the state-of-the-art Rust verifier Prusti. To this end, we have introduced novel type-dependent contracts, which, combined with specification extension traits, allow specifying standard-library iterators without modifying the source code of the standard library itself.

%% file: ms.bbl
\begin{thebibliography}{10}
\providecommand{\url}[1]{\texttt{#1}}
\providecommand{\urlprefix}{URL }
\providecommand{\doi}[1]{https://doi.org/#1}

\bibitem{Prusti_2019}
Astrauskas, V., M{\"{u}}ller, P., Poli, F., Summers, A.J.: Leveraging {R}ust
  types for modular specification and verification. Proc. {ACM} Program. Lang.
  \textbf{3}({OOPSLA}),  147:1--147:30 (2019). \doi{10.1145/3360573}

\bibitem{AxiomProfiler_2019}
Becker, N., M\"uller, P., Summers, A.J.: The axiom profiler: Understanding and
  debugging {SMT} quantifier instantiations. In: Tools and Algorithms for the
  Construction and Analysis of Systems (TACAS). pp. 99--116. LNCS, Springer
  (2019)

\bibitem{Denis_Creusot_2022}
Denis, X., Jourdan, J.H., March{\'e}, C.: {Creusot: a Foundry for the Deductive
  Verication of Rust Programs}. In: International Conference on Formal
  Engineering Methods (ICFEM). LNCS, {Springer Verlag}, Madrid, Spain (2022),
  \url{https://hal.inria.fr/hal-03737878}

\bibitem{Filliatre_Why3_2013}
Filli{\^a}tre, J.C., Paskevich, A.: Why3--where programs meet provers. In:
  European Symposium on Programming (ESOP). pp. 125--128. Springer (2013)

\bibitem{SAVCBSProceedings_2006}
Jacobs, B., Cok, D., Weide, B., Bierhoff, K., Krishnaswami, N., et~al.:
  Proceedings of the 2006 Conference on Specification and Verification of
  Component-based Systems (SAVCBS). {ACM} Digital Library (2006)

\bibitem{Jacobs_Iterators_2006}
Jacobs, B., Piessens, F., Schulte, W.: {VC} generation for functional behavior
  and non-interference of iterators. In: Specification and Verification of
  Component-based Systems (SAVCBS). p.~67. {ACM} Press (2006).
  \doi{10.1145/1181195.1181209}

\bibitem{Jacobs_VeriFast_2011}
Jacobs, B., Smans, J., Philippaerts, P., Vogels, F., Penninckx, W., Piessens,
  F.: Verifast: A powerful, sound, predictable, fast verifier for {C} and
  {J}ava. In: NASA Formal Methods Symposium. pp. 41--55. Springer (2011)

\bibitem{RustBook_2018}
Klabnik, S., Nichols, C., contributors: {The Rust Programming Language}. Rust
  Community (2018), \url{https://doc.rust-lang.org/stable/book/}

\bibitem{Krishnaswami_Iterators_2006}
Krishnaswami, N.R.: Reasoning about iterators with separation logic. In:
  Specification and Verification of Component-based Systems (SAVCBS). p.~83.
  {ACM} Press (2006). \doi{10.1145/1181195.1181213}

\bibitem{Rust_2014}
Matsakis, N.D., Klock, F.S.: {The Rust language}. ACM SIGAda Ada Letters
  \textbf{34}(3),  103--104 (Nov 2014). \doi{10.1145/2692956.2663188}

\bibitem{Viper_2016}
M{\"u}ller, P., Schwerhoff, M., Summers, A.J.: Viper: A verification
  infrastructure for permission-based reasoning. In: Verification, Model
  checking, and Abstract interpretation (VMCAI). pp. 41--62. Springer (2016)

\bibitem{Parkinson_SLAbstraction_2005}
Parkinson, M.J., Bierman, G.M.: Separation logic and abstraction. In: Palsberg,
  J., Abadi, M. (eds.) Principles of Programming Languages ({POPL}). pp.
  247--258. {ACM} (2005)

\bibitem{Pereira_Iterators_2018}
Parreira~Pereira, M.J.: {Tools and Techniques for the Verification of Modular
  Stateful Code}. Theses, {Universit{\'e} Paris Saclay} (Dec 2018),
  \url{https://tel.archives-ouvertes.fr/tel-01980343}

\bibitem{Reynolds_SL_2002}
Reynolds, J.C.: Separation logic: A logic for shared mutable data structures.
  In: Logic in Computer Science (LICS). pp. 55--74. IEEE (2002)

\bibitem{Smans_Snapshots_2010}
Smans, J., Jacobs, B., Piessens, F.: Heap-dependent expressions in separation
  logic. In: Formal Techniques for Distributed Systems, pp. 170--185. Springer
  (2010)

\bibitem{Wolff_Closures_2021}
Wolff, F., B{\'i}l{\'y}, A., Matheja, C., M{\"u}ller, P., Summers, A.J.:
  Modular specification and verification of closures in {Rust}. Proceedings of
  the ACM on Programming Languages  \textbf{5}(OOPSLA),  1--29 (Oct 2021).
  \doi{10.1145/3485522}

\end{thebibliography}
